\documentclass{iaus}
\usepackage{graphicx}
\usepackage{epsfig}

\title[Discovery of the magnetic field in the B1/B2V star {$\boldmath \sigma$} Lup] 
{The magnetic field of the B1/B2V star $\sigma$ Lup}

\author[Henrichs, Kolenberg, Plaggenborg, Marsden, Waite, Wade et al.]   
{H.F. Henrichs$^1$,
K. Kolenberg$^2$, B. Plaggenborg$^1$, S.C. Marsden$^3$,\\ I.A. Waite$^4$,
G. Wade$^5$, and the MiMeS Collaboration$^6$}

\affiliation{$^1$Astronomical Institute Anton Pannekoek, University of Amsterdam, Science Park 904, 1098XH Amsterdam, Netherlands, email: {\tt h.f.henrichs@uva.nl},\\
[\affilskip]
$^2$Institut f\"{u}r Astronomie, Universit\"{a}t Wien, T\"{u}rkenschanzstrasse 17, A-1180 Vienna, Austria \\
[\affilskip]
$^3$Anglo-Australian Observatory, PO Box 296, Epping, NSW 1710, Australia\\
[\affilskip]
$^4$Faculty of Sciences, University of Southern Queensland, Toowoomba, Qld 4350 Australia \\
[\affilskip]
$^5$Dept. of Physics, Royal Military College of Canada, Kingston, Canada\\ 
[\affilskip]
$^6$http:$//$www.physics.queensu.ca$/$$\sim$wade/mimes/MiMeS$\_$$\_$Magnetism$\_$in$\_$Massive$\_$Stars.html
}
\pubyear{2011}
\volume{272}  
\pagerange{100--100}
\date{August 2010 and in revised form ??}
\jname{Active OB stars - structure, evolution, mass loss, and critical limits}
\editors{C Neiner et al., eds.}
\graphicspath{{converted_graphics/}}
\begin{document}

\maketitle

\begin{abstract} The ultraviolet stellar wind lines of the photometrically periodic variable early B-type star $\sigma$ Lupi were found
to behave very similarly to what has been observed in known magnetic B stars, although no periodicity
could be determined.  AAT spectropolarimetric measurements with SEMPOL were obtained.  We detected a
longitudinal magnetic field with varying strength and amplitude of about 100 G with error bars of
typically 20 G. This type of variability supports an oblique magnetic rotator model. We fold the
equivalent width of the 4 usable UV spectra in phase with the well-known photometric period of 3.019
days, which we identify with the rotation period of the star. The magnetic field variations are
consistent with this period. Additional observations with ESPaDOnS attached to the CFHT strongly
confirmed this discovery, and allowed to determine a precise magnetic period.  Like in the other
magnetic B stars the wind emission likely originates in the magnetic equatorial plane, with maximum emission
occurring when a magnetic pole points towards the Earth.  The 3.0182 d magnetic rotation period is
consistent with the photometric period, with maximum light corresponding to maximum magnetic field. No
helium or other chemical peculiarity is known for this object.



\end{abstract}

\firstsection
\section*{Introduction and Analysis}

In nearly all magnetic OB stars the dipole component is dominant. As the rotation and magnetic axis do
not coincide in general, these objects act as oblique rotators.  The outflowing stellar wind is
perturbed by the surface magnetic field, and is periodically modified. In fact, the discovery of a
number of magnetic early-type stars was preceded by the discovery of strictly periodic wind
variability as observed in the UV, which appeared to be the strongest indirect indicator for the
presence of a magnetic field. By this method three magnetic B stars have been found: $\beta$ Cep (\cite{henrichs00}),
$\zeta$ Cas, and V2052 Oph (\cite{neiner03a, neiner03b}), with rotation periods of 12 d, 5.4 d and 3.6 d, respectively.

In our search for stellar wind variability in the IUE archives, we found that the B1/B2V star
$\sigma$ Lup had variable UV wind lines (Fig.\,1a), similar to other magnetic B stars. This
prompted us to observe this star with SEMPOL at the AAT, with follow-up observations in the frame of the MiMeS
Collaboration. We describe here the discovery of the magnetic field and its further analysis. 

The main stellar parameters of $\sigma$ Lup are (\cite{Leven06}): $V=4.4$, $\log(L/L_{\odot}) = 3.76 \pm 0.06$, $M/M_{\odot} =9.0 \pm 0.5$, $R/R_{\odot} =4.8 \pm 0.5$,
$T_{\rm eff} = 23000 \pm 500$ K, $v$sin$i =80 \pm 14$ km s$^{-1}$, and a photometric period $P_{\rm phot} = 3.0186 \pm 0.0004$ d.


From the LSD spectra we computed the mean longitudinal field ($B_l$). 
The smallest error
bars are about 16 G.
The best fit of the function 
$B_l (t) = B_0  + B_{\rm max}  \cos(2\pi(t- t_0)/P)$ gives:
$B_0  = 7 \pm 5$ G, $B_{\rm max}  = 106 \pm 9$ G, 
$P = 3.01819 \pm 0.00033$ d, and 
$t_0  = $ JD $ 2455103.12 \pm 0.56$ with a reduced $\chi^2 = 1.0$. 
This function together with the data as a function of phase are plotted in Fig.\,1b (middle).
We identify the photometric period with the rotation period. From $v$sin$i$ and the estimated stellar radius follows  $i > 50^{\circ}$. The
magnetic tilt angle $\beta$ is then constrained by the observed ratio $B_{\rm max} / B_{\rm min} = \cos(i +\beta)/ \cos(i - \beta) =
- 1.14^{+0.27}_{-0.36}$, implying $\beta$  close to $90^{\circ}$.

\begin{figure}[htp]
\centering
\leftline{\includegraphics[bb=16 0 560 825,height=6.6cm,keepaspectratio]{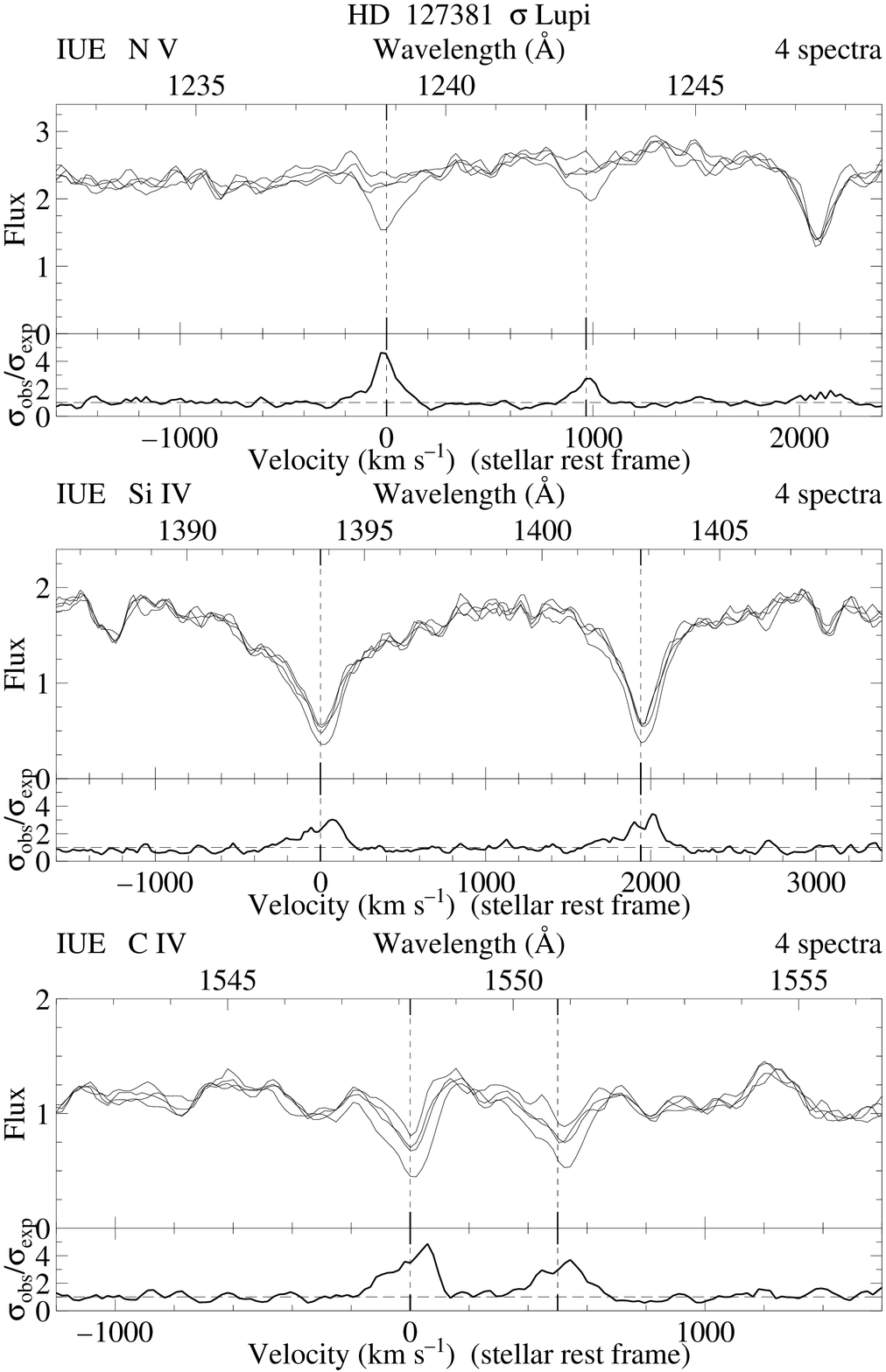}}
\hfill
\vspace{-6.9cm}
\rightline{\includegraphics[bb=18 164 498 789,height=6.6cm,keepaspectratio]{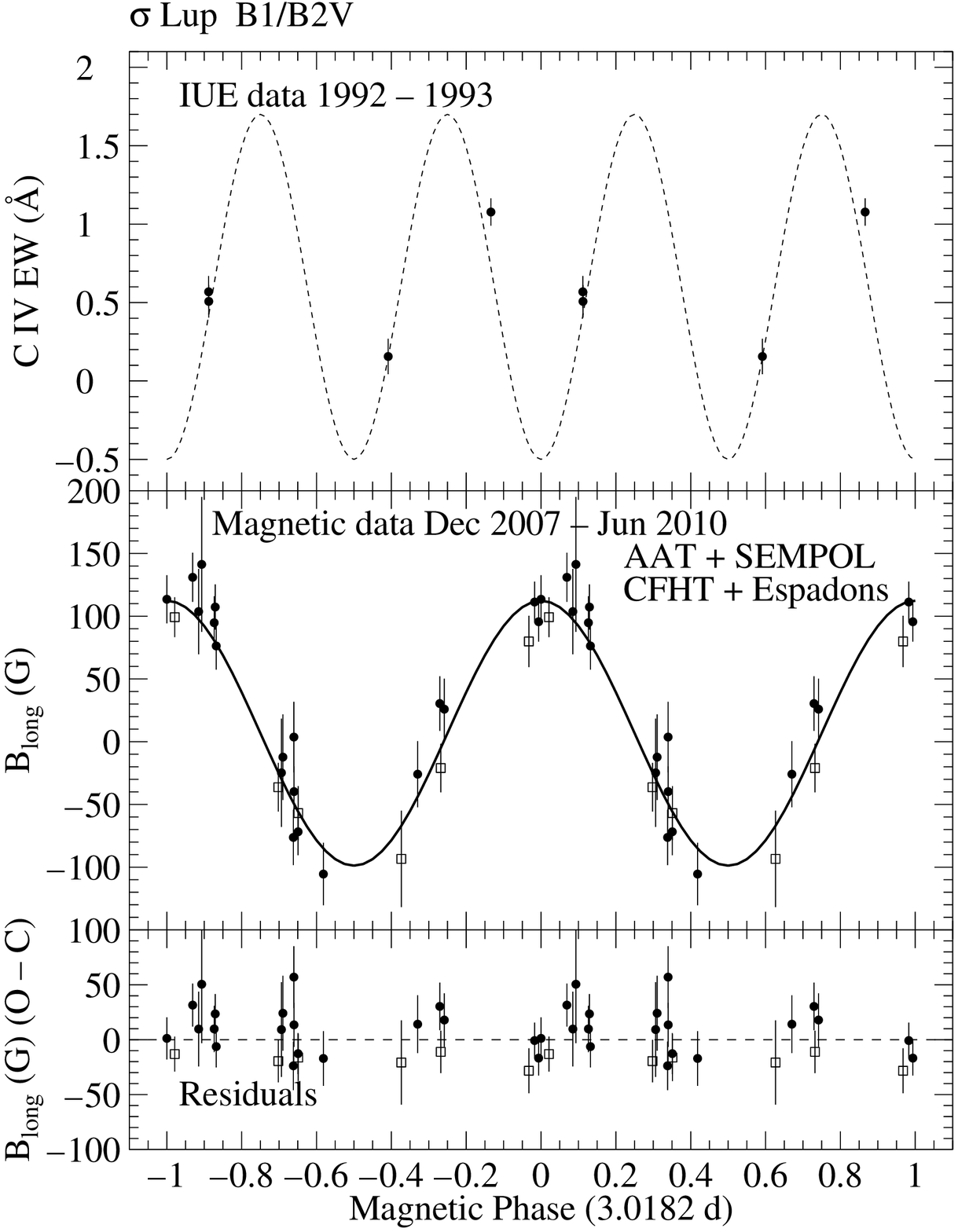}}
\caption{{\sl (a) Left:}  Variable UV wind lines of $\sigma$ Lup. Such variability is only found in magnetic B stars.
  {\sl (b) Right, middle:} phase plot of SEMPOL ($\square$) + ESPaDOnS ($\bullet$) magnetic data of $\sigma$ Lup with the weighted-fit period of
3.0182 d. The best cosine model fit is overplotted. {\sl Lower panel:} Residuals of magnetic data. {\sl Upper
panel:} Equivalent width of C IV wind lines. The expected double sine curve (not a fit), as observed in
other magnetic B stars is overplotted.
}
\end{figure}

The photometric period is $3.0186 \pm 0.0004$ d (\cite{Jerz92}), determined 3147 cycles earlier. 
The extrapolated epoch of maximum light coincides within the uncertainties with the epoch of maximum (positive) magnetic field.
The assumption that these epochs are equal allows a more accurate determination of the period: $P = 3.01858 \pm 0.00014$ d, i.e. within 12 s. New photometry would be able to confirm this.

The three previously discovered magnetic B stars ($\beta$ Cep, $\zeta$ Cas and V2052 Oph) showed a double sine
wave in the equivalent width of the UV wind lines, with the maximum emission (minimum EW) coinciding
with maximum field strength, i.e.\ at the phase when a magnetic pole is pointed towards the observer.
Because of the particular spacing of the 4 datapoints a fit of a double sine wave is not meaningful, but the expected curve (with arbitrary
scaling, see Fig.\,1b, top) suggests similar behavior for $\sigma$ Lup as well.
This supports a model with the emitting material in the magnetic equatorial plane.

\end{document}